\documentstyle[aps]{revtex}
\begin{document}

\pagestyle{myheadings}

\title{Spinodal decomposition in fluids: diffusive, viscous and
inertial regimes}

\author{Turab Lookman$^{1}$, Yanan Wu$^{1}$, Francis J.
Alexander$^{2}$
and Shiyi Chen$^{3,}$\footnote{On leave from Theoretical Division,
 Los Alamos National
Laboratory, Los Alamos, NM 87544}
}
\address{
${}^1$ Department of Applied Mathematics,\
University of Western Ontario,\
London, Ontario, Canada, N6A 5B7\\
${}^2$ Center for Computational Science,\
Boston University,\
3 Cummington Street,\
Boston MA 02215\\
${}^3$ IBM Research Division,\
T.J. Watson Research Center,\
P.O. Box 218,\
Yorktown Heights, NY 10598}

\date{submitted to Phys. Rev. Lett., July 19, 1995}
\maketitle

\begin{abstract}

Using a Langevin description of spinodal decomposition in fluids,
we examine
domain growth in the diffusive, viscous and
inertial regimes.  In the framework of this model, numerical
results  corroborate earlier theoretical predictions based on scaling
arguments
and dimensional analysis.

\end{abstract}

\vspace{0.3in}

The dynamics of phase transitions in binary fluids quenched
into the coexistence region has been the subject of
considerable study in recent years \cite{rev,bray}.
It is generally accepted that long
after the quench, the phase separation
dynamics can be characterized
by a single time dependent length scale, $R(t)
\sim t^{\alpha}$. As a result, much attention has
focused on how domains grow in time --- specifically
what is the growth exponent $\alpha$?

Scaling and dimensional analyses due to Siggia \cite{sig}, Furukawa
\cite{fur1},
San Miguel {\it et al} \cite{migu} and more recently Bray
\cite{bray} address this question.
 Experimental \cite{wong} and numerical
\cite{fv,shino,syctl,frank,wu,puri,koga,lsr}
studies, however, have not necessarily  supported  these theories,
sometimes providing conflicting results \cite{veltax1}.
Often overlooked
in spinodal decomposition in binary fluids is that
several stages of growth can occur, in each of which
a different transport mechanism dominates. This fact has been
remphasized in \cite{bray,fur1,bastea1,bastea2}.  Individual
experiments and numerical
simulations typically access only a particular regime.
Lacking has been a clear demonstration of 1) the existence
of these distinct regimes
within a single model and, subsequently, 2)
quantitative results in these regimes which validate theoretical
predictions \cite{bray,sig,fur1,migu}.
In this Letter we address these points.

To simulate phase separation in a binary fluid, we used the
Langevin model of Farrell and Valls
\cite{fv}.  The order parameter $\psi$ is the difference in the
concentration of the
two fluid components.
Its evolution and that of the fluid velocity are given by
\begin{equation}
\partial_{t} \psi = \Gamma \nabla^2 \mu - \lambda \nabla \cdot [
\psi \rho
{\bf u}]
\label{eq:order}
\end{equation}
\begin{equation}
\rho \partial_{t} {\bf u} = \eta \nabla^2  {\bf u} + \sigma \nabla
(\nabla
\cdot  {\bf u}) - \lambda \nabla \cdot  (\rho {\bf u} {\bf u}) - \lambda
\psi
\nabla
\mu
\label{eq:velo}
\end{equation}
where $\rho$ is the average mass density, $\Gamma$ is an order
parameter
diffusion coefficient, and $\eta$ is the shear viscosity. Here
$\sigma =\eta (1 - 2/d) + \zeta$, where $\zeta$ is the bulk
viscosity, and $d$ is
the spatial dimension. The dimensionless constant, $\lambda$,
couples the order parameter to the fluid velocity and is also the
strength of the convective flow. The
chemical
potential, $\mu=\frac{\delta F}{\delta \psi}$, where $F$ is the
free energy of the system at equilibrium given by
$F[\psi,{\bf u}] = \frac{1}{2} \int d^{d}{\bf r}[ \rho u^{2} +
\frac{1}{2}a\psi^{4} - b \psi^{2} + \beta K |\nabla \psi| ^2 ]$.
The strength of the interfacial energy is $\beta K$.
 Below the critical temperature, $a$ and $b$ are positive constants.

After the fluid is quenched, single phase droplets form and
grow. In the coarsening process a competition between
hydrodynamic and thermodynamic
effects can lead to three dynamical regimes: the diffusive, viscous and
inertial \cite {bray,fur1}. We discuss
these briefly, using dimensional analysis based on Equations
(1) and (2).

In the diffusive regime, the fluid velocities are small,  and
the advective term in (\ref{eq:order})
is negligible compared to the order parameter diffusion.
Therefore, (1) becomes $\partial_{t} \psi \sim \Gamma \nabla^2 \mu$.  Since
the chemical potential $\mu \sim {\kappa}/{R}$  where
$\kappa$ is the surface tension and $R$ is the characteristic length
scale
in the system (i.e. domain size), we have
$R(t) \sim (\Gamma \kappa)^{\frac{1}{3}} t^{\frac{1}{3}}$.
The coefficient $\Gamma \kappa$ implies that the growth in this
regime is
driven by diffusion and surface tension. In two dimensions,
for example, the surface tension $\kappa$ is given by
$\kappa
=\frac{4}{3}( 2 \beta K)^{1/2}$ \cite {fur1}.

In the viscous regime,  hydrodynamics becomes relevant. In
particular,
in the velocity equation (\ref{eq:velo}), the viscous term
dominates
the inertial terms.
If one ignores  the inertial and bulk viscosity terms, the shear
stress term
is balanced by the force due to the gradient in the chemical
potential. Thus,
$\eta \nabla^2 {\bf u} \sim \lambda \psi \nabla \mu$
so that
$R(t) \sim \frac{\lambda \kappa}{\eta} t$.
This is the linear growth law predicted by Siggia
\cite{sig}. The coefficient
$\frac{\lambda \kappa}{\eta}$ indicates growth driven by
the surface tension and controlled by the viscous force in the fluid.
The length-scale, $R_d$, and time-scale, $t_d$, at which
the system crosses over from the diffusion regime to the viscous
regime
is given by setting
$(\Gamma \kappa)^{\frac{1}{3}} t_{d}^{\frac{1}{3}} \sim
\frac{\lambda \kappa}{\eta} t_{d}$.
Thus,
 $t_d \sim (\frac{\Gamma \eta^3}{\kappa^2
\lambda^3})^{\frac{1}{2}}$
and
$R_d \sim (\frac{\Gamma \eta}{\lambda})^{\frac{1}{2}}$.

In the inertial regime,  inertial effects dominate over the viscous
forces so
that
$\rho \frac{d {\bf u}}{dt} \sim \lambda \psi \nabla \mu$.
This leads to
$R(t) \sim (\frac{\lambda \kappa}{\rho})^{\frac{1}{3}}
t^{\frac{2}{3}}$,
as predicted by Furukawa \cite{fur1} (see also \cite{bray}).
The coefficient here indicates that the growth is
driven by the surface tension and controlled by the inertial
effects.
The crossover between the viscous regime and the inertial regime
thus
occurs
at  length scale $R_h$ and time-scale $t_h$ where
$\frac{\lambda \kappa}{\eta} t_{h} \sim
(\frac{\lambda \kappa}{\rho})^{\frac{1}{3}} t_{h}^{\frac{2}{3}}$
so that $t_h \sim \frac{\eta^3}{\rho \lambda^2 \kappa^2}$ and
$R_h \sim
\frac {\eta^{2}}{\lambda \rho \kappa}$.
Similarly, the length, $R_i$, and time, $t_i$ for crossover from
diffusion directly
 to inertial
are given by $R_i \sim (\frac{\Gamma^2 \kappa \rho }{ \lambda} )^{\frac{1}
{3}}$
and
$t_i \sim \frac{\Gamma\rho}{\lambda}$ respectively. This would
correspond to the
inviscid flow case.

To facilitate growth of domains
in each of these regimes and to
access each of them within the framework of a single
model, we vary $R_d$ and $R_h$
(relative to system size) by
adjusting the
parameters
 $\eta$, $\lambda$ and $\beta$.
 For convenience, the actual (dimensionless)
numerical equations we solve are the
following:
\begin{equation}
\partial_{t} \phi = \nabla^2 [ \phi^3- \phi - \beta \nabla^2 \phi ] -
\ \hat{\lambda} \nabla \cdot [ \phi {\bf v}] + \ {\mu},
\end{equation}
\begin{equation}
\partial_{t} v_{i} = \ \hat{\eta} \nabla^2 v_{i} + \ \hat{\sigma}
\sum_{k}
\nabla_{i} \nabla_k v_k - \ \hat{\lambda} \phi \nabla_{i} [ \phi^3 -
\phi -
\beta\nabla^2 \phi ] - \ \hat{\lambda} \sum_{k} [ \nabla_k (v_i v_k)
+ v_k \nabla_i v_k ] + \ {w_i}.
\end{equation}
The rescaled order parameter and transport quantities are
given in terms of those used in (1) and (2) by
$\phi=(\frac{a}{b})^{\frac{1}{2}} \psi$, ${\bf v}=(\frac{a}{b \rho K})
^{\frac{1}{2}} {\bf
u}$,
$\hat{\eta}=\frac{\eta}{\rho \Gamma K}$,
$\hat{\sigma}=\frac{\sigma}{\rho \Gamma K}$, $\hat{\lambda}=\lambda
(\frac{b}
{\Gamma^2 a \rho K})^{\frac{1}{2}}$.
Space and time are rescaled by
${\bf r}\rightarrow{\bf r}$, $t\rightarrow\Gamma K t$. The dimensionless
crossover
lengths are
given by
$R_d=(\frac{\hat{\eta}}{\hat{\lambda}})^{\frac{1}{2}}$,
and
$R_h=\frac{\hat{\eta}^2}{\hat\lambda \hat\kappa}$,
where we have set  $\rho=1$. In two dimensions, the surface tension,
$\hat{\kappa}=\frac{4}{3} (2 \beta)^{\frac{1}{2}}$,
so that by varying $\beta$, we can control the surface tension.

We studied deep, critical or symmetric quenches with $\langle \phi
\rangle = 0$ throughout the
course of the simulations, where $ \langle . \rangle $ denotes
 an ensemble average or space
average.
 The order parameter and velocity are initially taken as
Gaussian
fields with $\langle \phi \rangle = \langle v_i \rangle =0$,
and $\langle \phi^2
\rangle = \langle v_i^2 \rangle = 0.005$.
The grid size $\Delta x$ used was 1.7 and the time step $\Delta t$
was chosen as 0.05 in two and 0.02 in three dimensions, respectively.
The numerical integration scheme is the same as  in \cite{fv,wu}.
The average domain size
was defined as  the first zero of the equal time correlation
function $G(r,t)=\langle \phi({\bf x},t)\phi({\bf x} + {\bf r} ,t) \rangle
$, the
Fourier transform of which is the structure factor,
$S(k,t)$.
The fields, $\mu$ and $w_i$, were Gaussian, white noise with
covariance given by
the fluctuation-dissipation relation\cite{fv,wu}. We  found
that adding
 noise does
not alter the growth exponent in the scaling regime. However, it
introduces
 curvature in the early growth
so that longer times are required to reach the scaling regime.
The results we report here were obtained in the absence of noise
and in all cases were averaged over 3 or 4 independent runs.

In two dimensions, on a $1024^2$ system, we let $\hat{\eta}=1$,
$\hat{\lambda}
=1$ and
$\beta=1$. Thus, $R_d \sim R_h \sim 1$ (in lattice units) are both
small
 compared to the lattice size, $L = 1024$,  so
that for domain size $R(t) \gg R_h$,
the system will favor droplet growth in the inertial regime. The
data
represented by ($\Box$) in Fig.1  shows
that $R(t)$ has behavior consistent with $\alpha=2/3$. In order to
have
a
viscous regime, one requires $R_d \ll R(t) \ll R_h$. This is satisfied
by
 choosing, for example,
$\hat{\eta}=20$, $\hat{\lambda}=1$ and $\beta=1$ so that
 $R_h \sim 120$ and $R_d \sim 3$.
In Fig.1  the symbol
($\times$) shows the growth under these conditions. It is
consistent with
 $\alpha=1/2$
 growth over a time interval spanning about 1.5  decades.
The exponent of $\alpha=1/2$ in
two dimensions was predicted in
 \cite {migu}.
Since $R_h \sim 120$, the inertial force would not be expected to
influence the growth until at late times when $R(t)$ is comparable with
$R_h$.
 To indicate how the $\alpha=1/2$ growth could yield
to the $\alpha=2/3$ growth, we changed parameters to make $R_h$
smaller so that the crossover from the viscous regime to
the inertial regime can happen earlier.
The symbols  ($\Diamond$) and (+) in Figure 1 show data for $R_h \sim  30$
($\hat{\eta}=11$, $\hat{\lambda}=1$, $\beta=1$) and  $R_h \sim  7$
($\hat{\eta}=5$, $\hat{\lambda}=1$, $\beta=1$) respectively. As $R_h$
(and $R_d$) decreases, the data shows that the viscous growth
and  a later, faster inertial  growth  occurs progressively earlier.
Finite size effects and the need for very long times to see adequate
viscous and inertial growth
make quantitative analysis of growth in the crossover regimes difficult.
 The time evolution of the
Reynolds number, $Re$,
the ratio of inertial  to viscous effects, is
consistent with the behaviour of $R(t)$ as $R_h$ decreases.
Corresponding to the parameters for domain growth, the insert to
Figure 1 shows how $Re$ changes from its behavior in the viscous
regime, where $Re < 1$ and is essentially constant, to that in the
inertial regime where it  increases as $t^{1/3}$ \cite{fur1}.
 We find that for $\eta \geq 11$,
the system lies well within the viscous regime until the influence of inertial
flow at very late times.
The scaled correlation functions $G(\xi)$ and
$\xi^{2}G(\xi)$ \cite{shino} are shown in Figure 2,
 for
 $\hat{\eta}=20$, $\hat{\lambda}=1$,
 $\beta=1$, where $\xi=r \langle k \rangle$
and $\langle k \rangle=\frac {\int kS(k,t) dk} {\int S(k,t) dk}$.
 The data collapses well for several
times, indicating that
the $\alpha=1/2$ growth
 is in the scaling regime. The scaling behavior for $\alpha=2/3$
was shown in
\cite{wu}.

Using this model in two dimensions, we previously \cite{wu}
examined
the behavior of $\alpha$ as a
function of the coupling constant, $\hat{\lambda} (0<\hat{\lambda}<1)$,
  by fixing
$\hat{\eta}$, $\beta$ and $\rho$. For this one parameter system,
$R_h \sim
\frac{1}{2\hat{\lambda}}$ and $R_d \sim \frac{1}{\hat{\lambda}^{1/2}}$.
For  $1/2<\hat{\lambda}<1$, $R_h \sim R_d \sim 1$,
so that only the inertial growth survived.
 For $\hat{\lambda} \rightarrow 0$, the domain size $R(t) < R_d$
 and the dominant
mechanism was diffusion. It was the first attempt to show within a
single model different regimes, however,
the one--parameter system was limited in how well it could
capture all three regimes.
The competing mechanisms of viscosity, inertial force
and surface tension
appear to demand a system with two parameters.
Two--dimensional lattice Boltzmann and lattice gas simulations
seem to be carried out primarily with
relatively small $R_h$, thus the $\alpha=2/3$  estimates
are consistent with growth in the inertial regime \cite
{syctl,frank,bastea1}. The results from molecular dynamics are
controversial.
It has been pointed out \cite{bastea2} that the $\alpha=1/2$
growth obtained in
\cite{lsr}
may be attributed to droplet coalescence.
 Velasco and Toxvaerd \cite {veltax1} observed  $\alpha=1/2$
 crossing over to $\alpha=2/3$ in their two-dimensional molecular
dynamics
simulations.

Three-dimensional simulations were carried out on Equations (3) and (4)
using a system
with
$256^{3}$ lattice sites and show behavior
analagous
 to that observed in two
dimensions. As above, we set $\beta=1$.
If $\hat{\eta}=1$ and $\hat{\lambda}=1$, $R_h \sim R_d \sim 1$ and
 are small compared
 to the
domain
size $R(t)$. One thus expects inertial growth with $\alpha = 2/3$
at late
 times, and this
is seen by the data represented by ($\Box$) in
Fig. 3.
If $\hat{\eta}=\hat{\lambda}$ (with $R_d \sim 1$), the system should favor
 growth in the viscous
regime for sufficiently large $\hat{\eta}$. The symbols ($\times$) and
($\Diamond$) in Fig.3 show growth for   $\hat{\eta}=25$,
$\hat{\lambda}=25$,
  and $\hat{\eta}=20$, $\hat{\lambda}=20$, respectively. As $\hat{\eta}$
increases, the growth  becomes consistent with $\alpha=1$.
 The crossover between the viscous regime and the
inertial
regime can be simulated through decreasing $R_h$, while keeping
$R_d$ small ($\sim 1$).
The symbols $(+)$ in Fig. 3 used $\hat{\eta}=12$,
$\hat{\lambda}=12$ and show that a regime with a growth
exponent of $1$
gradually yields to a slower growth regime, a $2/3$ type growth.
Figure 3 (insert) shows that the behaviour of the Reynolds number, Re,
is consistent with growth for  appropriate parameters in the inertial and
 viscous regimes.
 Finite size effects
are more pronounced in three dimensions, so that the inertial regime
is difficult to access as $R_h$ increases.
In Fig. 4 is plotted the scaling of the correlation functions
$G(\xi)$ and $\xi^2G(\xi)$ (insert)
 for the inertial regime in three dimensions.
The quality of the collapse of the data in Fig. 4 for several times
indicates that the $\alpha=2/3$ growth is in
the
scaling regime.

 Earlier work by Farrell and Valls \cite{fv} on the same model
was carried out on an $81^{3}$ lattice with $\hat{\eta} \sim 1$,
$\hat{\lambda} \sim 1$,
 and $\hat{\sigma}=2$ so that $R_h\sim 1$. Their
estimate of $\alpha \sim 1$ was based on an extrapolation
of a time-dependent,  effective exponent in terms of inverse droplet size.
Puri and Dunweg \cite{puri} used a Cell Dynamical System model
and
obtained
$\alpha \sim 1$ on a model
 (with $80^{3}$ lattices)
without the convective term in the velocity equation and with
$\hat{\eta}=1$,
$\hat{\lambda}=2$ ,and $\hat{\sigma}=2$. Using their
parameter values
{\em with} the convective term on a $128^3$ lattice,  we find
 an early $\alpha \sim 1$ growth
that crosses over to a slower $\alpha \sim 2/3$ growth at later times.
Shinozaki and Oono\cite{shino} and Koga and
Kawasaki\cite{koga}  obtained
 $\alpha \sim 1$ at late times with their models (Model H), ignoring
the inertial terms. It was noted in \cite{shino} that for larger
values of viscosity there is a crossover from $\alpha \sim 1/3$ to
$\alpha \sim 1$ growth. Such a crossover can occur because a
 larger viscosity increases $R_d$ which then
favors the diffusive
growth
for domain sizes $R(t) < R_d$.
Lattice Boltzmann simulations also provide linear growth estimates
\cite{syctl,frank}.
The model we have used allows for slight compressibility. However,
it has been shown that in the viscous regime $\alpha$ does not
change with the incompressible condition \cite{shino,koga}.
To our knowledge, the $\alpha=2/3$ in the inertial regime has not
been
observed in experiment or three-dimensional
simulations.

In summary, we have used a single model system to probe
the hydrodynamic regimes that a phase separating fluid can
undergo. In
particular, we have shown how domain growth can be favored to
take
place in these
regimes by an appropriate choice of the crossover lengths
$R_d$ and $R_h$ within a finite size simulation. Moreover, we
have obtained values for the growth exponent $\alpha$ in these
regimes
 in two and three dimensions that are in agreement with the
predictions of
 scaling
 and dimensional arguments. Our work helps to explain the
estimates of growth exponents, $\alpha$, obtained in a number of previous
studies.

We thank  B. J. Alder, S. Bastea, R. Desai and J. L. Lebowitz for
helpful discussions.
T.L acknowledges support from the National Science and
Engineering Research Council of Canada. Y.Wu is grateful to the
University of Western
Ontario for a graduate research fellowship.
Numerical simulations were carried
out using the computational resources of the Advanced Computing
Laboratory
at the Los Alamos National Laboratory and the Supercomputing
Center at
the University of Minnesota.

\newpage

\noindent {\bf Figure Captions}

\begin{description}

\item{FIG. 1.} The domain growth $R(t)$ vs time $t$ in {\em two}
dimensions
for different crossover lengths $R_h$ showing the change from growth
in the inertial regime ($t^{2/3}$) to the viscous regime ($t^{1/2}$).
 The errors in the data due to different initial conditions
 are of the order of the size
of the symbols.
The insert shows the Reynolds number, $Re=\tilde{v}R(t)/\hat{\eta}$, as a
function of time, where $\tilde{v}$ is the characteristic velocity
calculated as $dR/dt$ \cite{fur1}.
 The data are consistent with  $Re \sim t^{1/3}$
in the inertial regime and $Re \sim constant$ in the viscous regime.
 The straight line has slope $1/3$.
 The symbols represent
$\hat{\eta}=20,\hat{\lambda}=1,
R_h \sim 120 (\times)$, $\hat{\eta}=11,\hat{\lambda}=1,
R_h \sim 30 (\Diamond)$,
$\hat{\eta}=5,\hat{\lambda}=1,
R_h \sim 7 (+)$ and $\hat{\eta}=2,\hat{\lambda}=1,
R_h \sim 1  (\Box)$. The surface tension controlling parameter
$\beta=1$ for all cases.

\item{FIG.  2.} The scaled and normalised
 correlation function $G(\xi)$ vs $\xi$, where
$\xi=r \langle k \rangle$ and the first moment
 $\langle k \rangle=\frac{\int k S(k,t) dk}{\int
S(k,t)}$,
for $\hat{\eta}=20$, $\hat{\lambda}=1$, and $\beta=1$ on $1024
\times 1024$
lattices, at $t=2,500$ ($\Diamond$), $3000$ ($\times$), $4000$
($\circ$), $5,000$
(+) and $6,000$ ($\Box$). The insert shows $\xi^{2}G(\xi)$ versus $\xi$
for the same times
 and since
$ \langle k \rangle \sim 25$, scaling is good for $\sim 80$ lattice
units.

\item{FIG. 3.}
The domain growth $R(t)$ vs time $t$ in {\em three} dimensions
for values of crossover lengths $R_h$ showing the change in $\alpha$
from inertial ($t^{2/3}$) to viscous ($t$) regime.
The errors are of the size of the symbols.
 The insert shows that the behaviour of the Reynolds number,
$Re$, is consistent with $t^{1/3}$ in the inertial regime and $t$ in the
viscous regime. The straight lines have slopes $1/3$ (-- - --) and
 $1$ (-----).
The symbols represent
$\hat{\eta}=25,\hat{\lambda}=25,
  (\circ)$,
$\hat{\eta}=20,\hat{\lambda}=20,
  (\Diamond)$,
 $\hat{\eta}=8,\hat{\lambda}=8,
  (+)$,
 $\hat{\eta}=1,\hat{\lambda}=1,
  (\Box)$. The surface tension controlling parameter
$\beta=1$ for all cases.

\item{FIG. 4.}
The scaled and normalised correlation function $G(\xi)$ vs $\xi$
  for
 $\hat{\eta}=\hat{\lambda}=1$ and $\beta=1$ on $256^3$
lattices,
at $t=600$ ($\diamond$), $800$ ($\Box$), $1,000$ ($\times$),
and $1,200$ ($\circ$).
The insert shows $\xi^2 G(\xi)$ vs $\xi$
for the same times. Since $\langle k \rangle \sim 6$, the times scale well to
$\sim 40$ lattice units.

\end{description}
\end{document}